\documentclass{aastex701}
\usepackage{graphicx}
\usepackage{booktabs}
\usepackage{amsmath}
\usepackage{siunitx}
\usepackage{gensymb}
\usepackage{float}
\usepackage{caption}
\usepackage{subcaption}

\usepackage{xcolor}

\newcommand{\fxta}{FXT-A}
\newcommand{\fxtb}{FXT-B}
\newcommand{\ep}{Einstein Probe}
\newcommand{\erosita}{eROSITA}

\begin{document}

\title{{IC 1297: The Most Distant Known X-ray Planetary Nebula and One of the Most Luminous}}

\correspondingauthor{Martin A. Guerrero}
\author[0009-0005-6854-9619]{Haoyang Yuan}
\affiliation{Laboratory for Space Research, The University of Hong Kong, Hong Kong, P.~R.~China}
\affiliation{Department of Physics, The University of Hong Kong, Hong Kong, P.~R.~China}
\affiliation{Department of Automation, Tsinghua University, Beijing, P.~R.~China.}
\email{yuanaa@connect.hku.hk}

\author[0000-0002-2062-0173]{Quentin Parker}
\affiliation{Laboratory for Space Research, The University of Hong Kong, Hong Kong, P.~R.~China}
\email{quentinp@hku.hk}

\author[0000-0002-7759-106X]{Martin A.\ Guerrero}
\affiliation{
Instituto de Astrof\'\i sica de Andaluc\'\i a, IAA-CSIC, 
Glorieta de la Astronom\'\i a s/n, 
Granada E-1800, Spain
}
\email[show]{mar@iaa.es}

\begin{abstract}
We present the first targeted Einstein Probe (EP) X-ray observations of any planetary nebula (PN).
Using the onboard Follow-up X-ray Telescope (FXT), we detect
X-ray emission within the central region of  IC\,1297. 
The spectral analysis yields a plasma temperature of $kT = 
0.086^{+0.054}_{-0.058}$~keV ($T\sim1.0 \times 10^{6}$~K), characteristic of shocked wind bubbles in PNe, whereas there is no evidence that its Wolf-Rayet central 
star emits X-rays.
The absorption-corrected X-ray luminosity in the 0.3--2.0 keV energy band is $L_{\mathrm{X}} = 2.8 \times 10^{32}$ 
erg~s$^{-1}$, also consistent with a hot bubble. 
At an adopted distance of $\sim3.8~kpc$, IC\,1297 is the most distant X‑ray emitting PN found to date, opening up a new observational 
window on hot-bubble formation and evolution. It is also among the most luminous known X-ray-emitting PNe.
Our results demonstrate the capability of the EP FXT for detecting and characterizing faint, 
extended X-ray sources in the soft X-ray band for distant PNe, extending the known spectral properties of detectable X-ray emitting PNe.

{We also compiled the X-ray properties of all 41 known Galactic
XPNe, finding a clear correlation between distance and X-ray
luminosity that reflects an observational bias toward detecting
the brightest sources at large distances. The compilation also
shows a separation in X-ray luminosity between hot bubbles powered
by H-rich and [WR]-type central stars, with the latter reaching
luminosities up to an order of magnitude higher. The ten most
luminous XPNe all host [WR]/PG\,1159 central stars.}


\end{abstract}


\section{Introduction}
\label{sec:introduction}
Planetary nebulae (PNe) are the ionized ejecta of low- to intermediate-mass stars ($0.8$--$8\,M_{\odot}$) 
during the final stages of their evolution, marking the transition from the asymptotic giant
branch (AGB) to the white-dwarf cooling sequence \citep[e.g.,][]{Kwok1978,Kwitter2022}.
As the stellar core is exposed, it drives a fast stellar wind ($v_{\infty}\gtrsim10^{3}$~km~s$^{-1}$) 
that overtakes the dense, slow-moving AGB envelope ($v_{\mathrm{AGB}}\sim10$~km~s$^{-1}$). 
In this interacting stellar winds (ISW) model \citep{Kwok1978}, the supersonic collision between these two winds generates a reverse shock that propagates inward into 
the fast wind, heating the post-shock material to temperatures that can exceed $10^7$~K 
\citep{Weaver1977}. This shock-heated, overpressured region is confined within the bright inner 
rim of the nebula and is commonly referred to as the ``hot bubble''. It constitutes the only extended component of the 
PN structure that can radiate efficiently at X-ray energies \citep[e.g.,][]{Kastner2000,Guerrero2000}.
X-ray observations of PNe are a key factor in assessing the action of fast stellar winds and collimated outflows 
on their formation and evolution. Any addition to the modest number ($\sim 40$) of X-ray-detected PNe (hereafter XPNe) is valuable.

{The first unambiguous detections of X-ray emission from PN
hot bubbles benefited from the arcsecond spatial resolution of
\textit{Chandra} and the sensitivity and collecting area of
\textit{XMM-Newton}. The moderate spectral resolution of the CCD
detectors on both observatories also helped distinguish soft
diffuse emission from harder point-like emission
\citep[e.g.,][]{Kastner2000,Chu2001,Guerrero2002}.} These observations confirmed 
the central prediction of the ISW model and revealed that the X-ray-emitting plasma is an 
optically thin gas with characteristic temperatures of \(T_{\mathrm{X}}\sim(1\text{--}3)\times10^{6}\)~K 
and luminosities of \(L_{\mathrm{X}}\sim10^{30}\text{--}10^{32}\)~erg~s$^{-1}$ in the \(0.3\text{--}1.5\)~keV 
band \citep[e.g.,][]{Ruiz2013,Toala2014,Toala2016}. These early studies established the basic physical conditions of hot bubbles and paved the way for later detailed studies of their formation and evolution. 
They also unveiled X-ray emission from their central stars (CSPNe) with energies above those expected from their photospheric emission \citep{Guerrero2001}. 

The \textit{Chandra} Planetary Nebula Survey \citep[ChanPlaNS;][]
{ChanPlaNS1,ChanPlaNS2,ChanPlaNS3} represents the most systematic census of XPNe to date. 
It comprised a representative sample of mostly high-excitation PNe, many of which are compact, within $\sim1.5$~kpc of the Sun \citep{ChanPlaNS1}. {The ChanPlaNS survey was subsequently extended to include additional compact PNe within the same approximate distance limit \citep{ChanPlaNS2}.} It was based largely 
on the volume-limited Galactic PN samples of \cite{2008PhDT.......109F}. Of the 59 PNe observed, roughly 70\% have X-ray 
detections with 30\% displaying diffuse hot-bubble emission \citep{ChanPlaNS2,ChanPlaNS1}. 
The largest share of X-ray-emitting PNe is due solely to point-source emission originating directly from their CSPNe. 
{Many CSPN point sources exhibit relatively hard X-ray
components at energies $\gtrsim0.5$~keV that are unlikely to
originate from the stellar photosphere. Within this point-source
population, \citet{ChanPlaNS3} identified a higher-temperature
group whose X-ray luminosities appear uncorrelated with the
CSPN bolometric luminosities, and a lower-temperature group with
$L_{\rm X}/L_{\rm bol}\sim10^{-7}$. They suggested magnetically
active binary companions and self-shocking stellar winds,
respectively, as possible origins of these two groups.}

Targeted PNe observations beyond the ChanPlaNS volume have been rare. 
The most distant XPN with a confirmed 
\textit{Chandra} hot-bubble detection is IC\,4593 at $\approx2.4$~kpc \citep[
][]{ic4593}, underscoring how little is known 
about hot-bubble properties at distances representative of the broader Galactic PN population. 
This observational gap is driven by the relatively weak and soft X-ray emission from PNe, such that 
geometric dilution with distance and extinction rapidly diminishes X-ray observability and strongly limits detections to local, 
high surface brightness examples.  
In general the evolutionary state, progenitor mass, central star properties of PNe, and the surrounding dust and 
ambient interstellar medium (ISM) density along PNe sight lines all vary systematically across the Galaxy to a greater 
degree than is present in the local sun centred volume. Consequently, the PNe located within such a volume 
are less representative of the population's overall diversity.

Indeed, all previous PNe X-ray observations have been hampered by facilities with poor, low-energy sensitivity 
\citep[e.g. ROSAT,][]{1999A&A...349..389V} or limited field of view (e.g.\ Chandra, XMM-Newton). This biases samples toward compact, high-surface-brightness, nearby PNe. The eROSITA All-Sky Survey \citep[eRASS,][]{2021A&A...647A...1P, 
2024A&A...682A..34M} opened a new window for identifying XPNe candidates across the sky \citep{Yuan_2025}.  
The \ep\ satellite \citep{2025SCPMA..6839501Y}, launched in 2024, allows deeper follow-up of these XPN candidates.


Here we report the detection of diffuse X-ray emission from IC\,1297, a well known, double ring PN at a distance of 3.8~kpc
\citep{FrewDistance} using observations with the Einstein Probe observatory. 
IC\,1297 is a small, southern-sky PN (Right ascension $19h~17m~23.45s$, Declination $-39^{\circ}~36^{\prime}~46.25^{\prime\prime}$, J2000). In this work, Gaia DR3 source 6714803311390242048 \citep{GaiaDR32023} and the optical CSPN of IC\,1297 are considered to be the same source. The PN was initially discovered by the Scottish 
astronomer Williamina Fleming in 1894. It has strong multi-wavelength 
radio, WISE MIR, 2MASS NIR, Optical and GALEX UV detections, as summarised in the HASH database \citep{HASH2016}, 
where it has HASH ID\#1267. The central star is classified as a hot Wolf–Rayet ([WR] hereafter) with spectral 
type [WC3/4] or [WO3], indicating fast stellar winds ($\sim 2,500~km/s$) and strong carbon and oxygen emission 
lines typical of very evolved stars \citep{2004A&A...427..231G}. Among the [WR]-type central stars in the ChanPlaNS sample, the diffuse hot bubble detection rate is very high, which makes them 
significantly over-represented compared to the hot bubbles of XPNe with standard, 
hydrogen-rich (O-type) central stars.

This work extends the known distance horizon for hot-bubble X-ray detections by a factor of $\sim1.54$, opening a new 
observational window on hot-bubble formation and evolution in the broader Galactic PN population that would benefit from further 
targeted observations. The remainder of this paper is organized as follows. Section \ref{sec:observations} describes the 
observations and data reduction procedures. Section \ref{sec:xray_properties} presents the spatial and spectral analysis methods, 
followed by discussion in Section \ref{sec:discussion} and conclusions in Section \ref{sec:conclusions}.

\section{X-ray Observations and Data Reduction}
\label{sec:observations}

\noindent

IC\,1297 was observed by the \ep\ observatory on August 2025 (obsID 11900356352, PI: H\,Yuan, co‑Is: M.\,A.\,Guerrero and Q.\,A.\,Parker). 
It carries the Follow-up X-ray Telescope modules \fxta\ and \fxtb\ that are two independent Wolter-I grazing-incidence X-ray 
telescopes operating in the 0.3--10 keV band. FXTB utilizes the eROSITA flight-spare mirror assembly, whereas FXTA employs an 
equivalent mirror assembly provided by ESA. They provide a large field of view ($1\arcdeg \times 1\arcdeg$) and moderate 
angular resolution (Half-Power Diameter HPD~$\sim22$\arcsec). 
With a combined effective area of $\simeq600$~cm$^2$ at 1~keV, the \ep\ FTX instrument reaches a sensitivity of $\sim$$10^{-14}$~erg~cm$^{-2}$~s$^{-1}$ in a 10~ks 
exposure \citep{Zhang_2025}. We note that \fxtb\ exhibits a slightly narrower PSF and a modestly larger effective 
area than \fxta, but the difference is at the $\sim10\%$ level, with both modules delivering comparable scientific performance.

The observations of IC\,1297 consisted of 18 separate targeted pointings, each observed simultaneously with both \fxta\ and \fxtb\ modules in the standard photon-counting 
mode, with a time resolution of 0.1~s. 

The effective exposure times accumulated by the two FXT modules are summarized in Table~\ref{tab:detection}. 
The total integration was 39.4~kiloseconds. Data processing was performed using the EP/FXT standard pipeline 
software (version 1.30) supplemented with custom specialised Python scripts written by the first author. 
\begin{deluxetable}{lccccc}
\tablewidth{0pt}
\tablecaption{Source Count Rates and Exposures\label{tab:detection}}
\tablehead{
\colhead{Module} &
\colhead{Effective Exposure} &
\colhead{Median Energy} &
\colhead{Counts} &
\colhead{Count Rate} \\
&
\colhead{(ks)} &
\colhead{(keV)} &
\colhead{} &
\colhead{(ks$^{-1}$)}
}
\startdata
EP-\fxta       & 19.706 & 0.576 & $37.2 \pm 7.8$   & $1.89 \pm 0.40$ \\
EP-\fxtb       & 19.677 & 0.564 & $127.8 \pm 12.7$ & $6.49 \pm 0.65$ \\
EP-\fxta+\fxtb & 39.382 & 0.566 & $165.0 \pm 14.9$ & $4.19 \pm 0.38$ \\
eROSITA        & 3.967  & 0.44  & $4.9 \pm 2.5$    & $1.24 \pm 0.62$ \\
\enddata
\tablecomments{Counts and count rates are background-subtracted using the source and background apertures shown in Fig.~\ref{fig:channels}. FXT counts are extracted in 0.3--10\,keV; eROSITA counts are extracted in 0.2--2.3\,keV. }
\end{deluxetable}

\begin{figure}[h]
\centering
\includegraphics[width=0.475\linewidth]{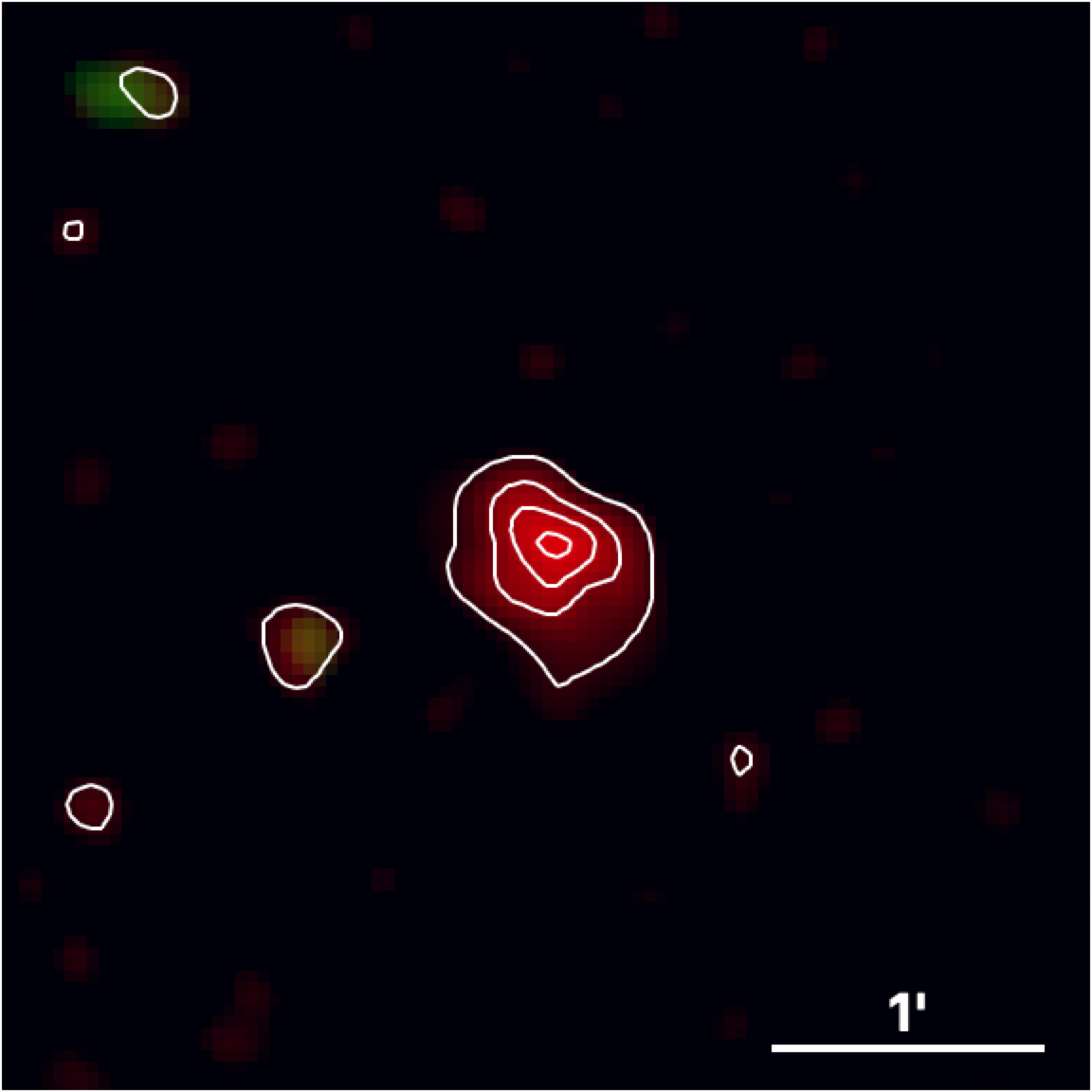}
\includegraphics[width=0.475\linewidth]{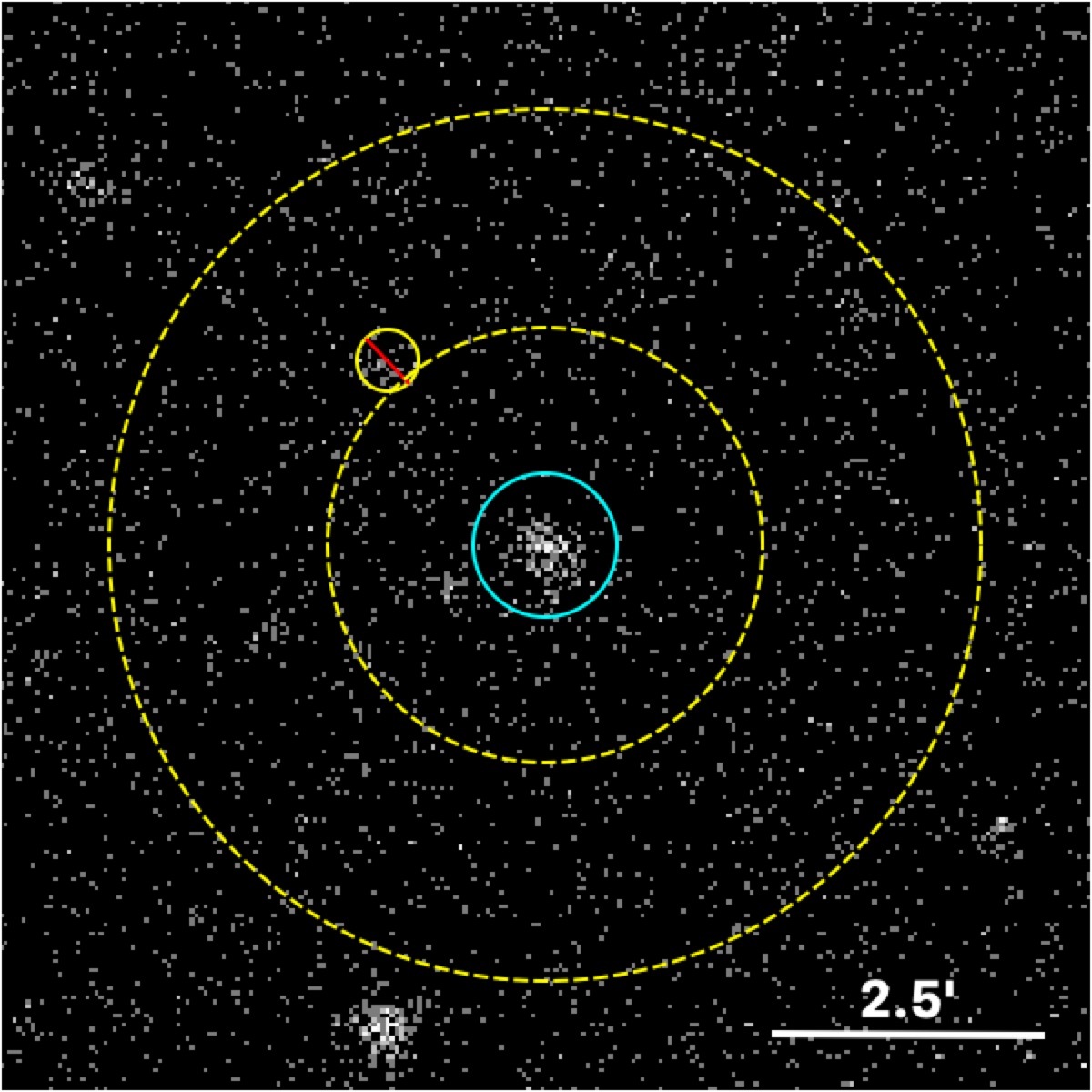}
\includegraphics[width=0.475\linewidth]{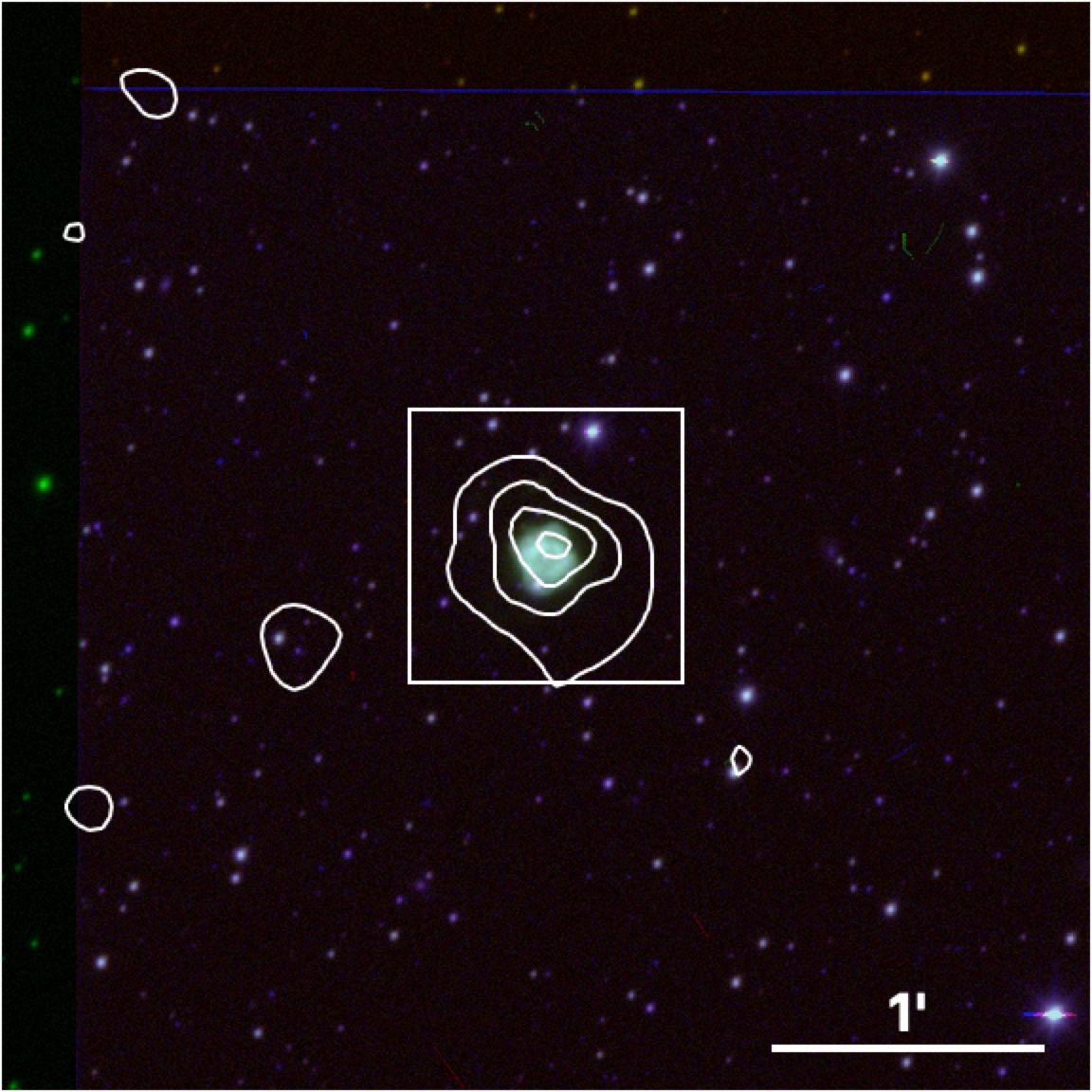}
\includegraphics[width=0.475\linewidth]{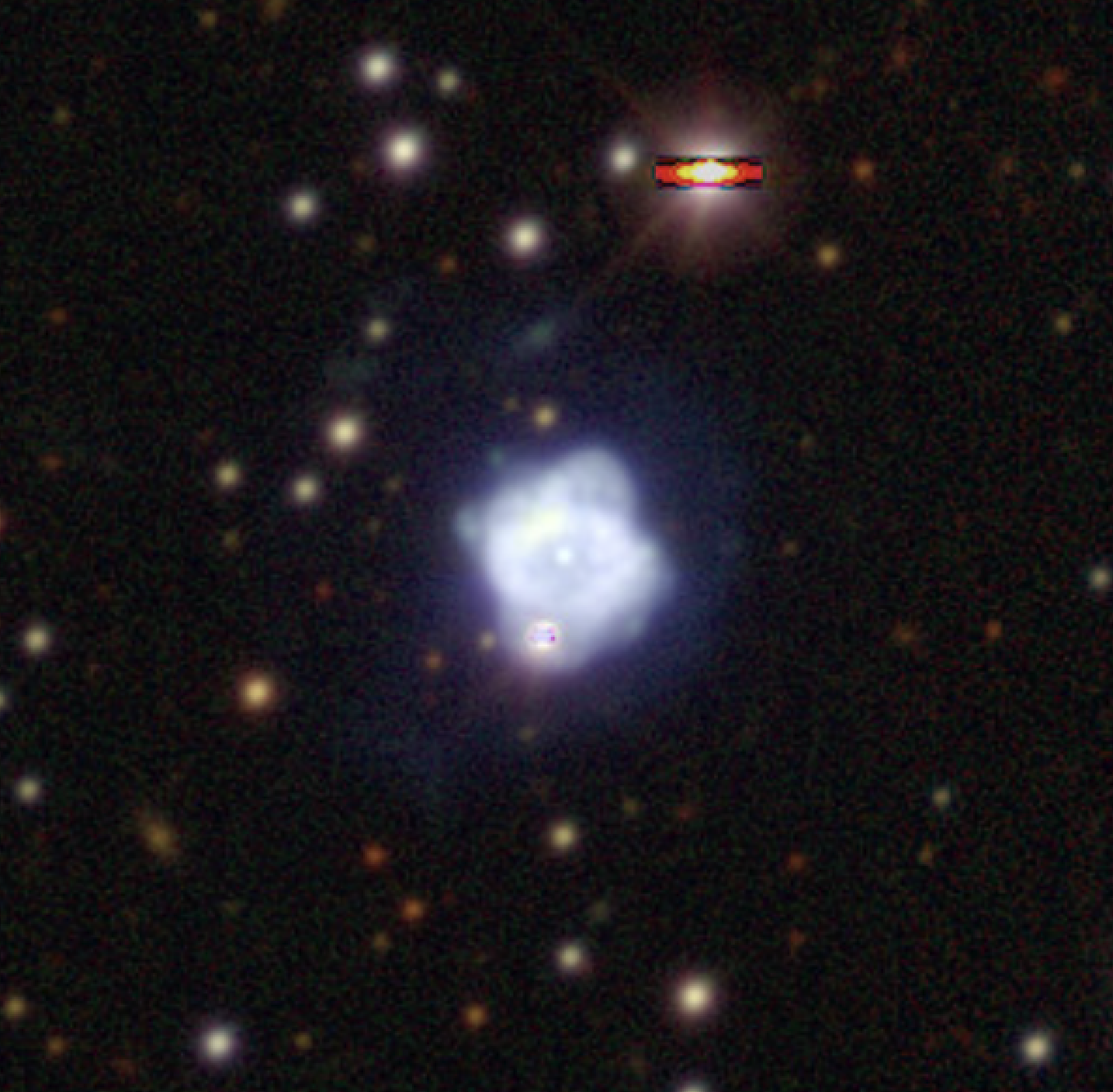}
\caption{
(top-left) \ep\ FXT RGB colour-composite picture of IC\,1297 in the soft 0.3-1.3 keV (red), medium 1.5-4.5 keV (green), and hard 5.0-10.0 keV (blue) energy bands. 
The X-ray contours correspond to the soft band and are spaced linearly. 
(top-right) \ep\ grey-scale image of IC\,1297 in the 0.3-10.0 keV band. 
The cyan circle corresponds to the source aperture used for spectral extraction. 
The background was extracted from the dashed yellow annular region, where a point source from the yellow solid circular aperture was excised. 
(bottom-left) DECAM 4m Blanco RGB colour-composite picture of IC\,1297 in the $r$ (red), $g$ (green), and $z$ (blue) optical bands. The X-ray contours of the soft band are overimposed. 
(bottom-right) $1'\times1'$ zoom of the bottom-left panel showing more detail including the inner fractured ring that has a prominent N-E enhancement, the multi-polar nature of the nebula and the [WR]
central star. A faint extended halo in the $g$ band surrounds the nebula. 
Image credit: Legacy Surveys DR10 (filters $g$, $r$, $z$)/ D.~Lang (Perimeter Institute) \& Meli thev, licensed under CC~BY~4.0, via Wikimedia Commons (cropped from the original). Available at: \url{https://commons.wikimedia.org/wiki/File:IC_1297_legacy_dr10.jpg}. North-East is to the top left in all panels.
}
\label{fig:channels}
\end{figure}

\section{X-ray Properties of IC\,1297}
\label{sec:xray_properties}

\subsection{X-ray Detection}

The EP observations clearly detect X-ray emission from IC\,1297, confirming the preliminary \erosita\ detection \citep{Yuan_2025}. 
To assess the spectral properties of this emission and its association with IC\,1297, images in the 
soft 0.3-1.3 keV, medium 1.5-4.5 keV, and hard 4.5-9.0 keV energy bands have been built with a pixel 
size of \SI{7.3}{\arcsecond}. This spatial binning, approximately one-third of the FXT's HPD of \SI{22}{\arcsecond}, 
balances angular resolution against signal-to-noise ratio. 
The X-ray emission in the top panels of Fig.~\ref{fig:channels} is confined to the soft band, 
with no detection at energies above 1.3 keV. 
The crude X-ray emission morphology appears elongated along a position angle of $\simeq 50^\circ$. The X-ray contours overlaid on the optical image (bottom-left panel of Fig.~\ref{fig:channels}) confirm that the detected emission originates from IC\,1297.

\subsection{Spectral Analysis}
\label{subsec:spectral_results}

A circular source aperture of radius $40\arcsec$, corresponding to approximately twice the optical 
PN nebular extent, was used to extract the X-ray emission from IC\,1297. The background was estimated 
from a concentric annulus with inner and outer radii of $2\arcmin$ and $4\arcmin$, respectively. Point 
sources within this background region were excised to minimise contamination (Fig.~\ref{fig:channels}, top-right). We use 
FXTDAS\citep{epfxt_fxtdas_2024} Version 1.30 to extract the corresponding spectra and calibration files.

\begin{figure}[h]
    \centering
    \includegraphics[width=0.48\columnwidth]{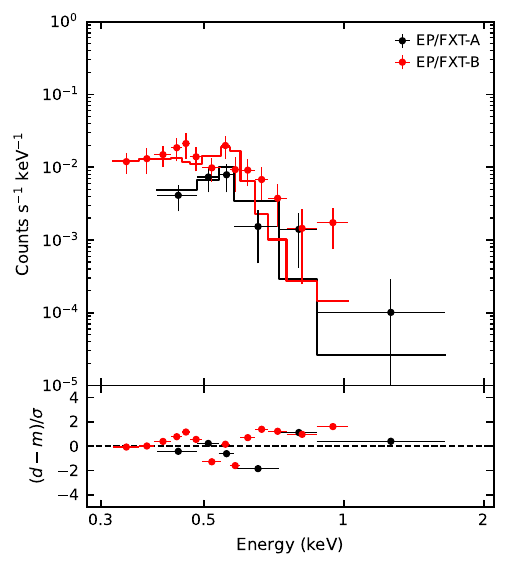}
    \caption{
        X-ray spectra of the target extracted from EP/FXT-A (black) and
EP/FXT-B (red) in the 0.3--2.0\,keV band.
The solid lines show the best-fitting absorbed thermal plasma model
(\texttt{tbabs}$\times$\texttt{apec}), with 
$N_{\rm H}=2.9\times10^{20}$\,cm$^{-2}$ and
$kT=0.086$\,keV. 
The bottom panel shows the fit residuals (data minus model) in units of $\sigma$.
Solar abundances are from \citet{wilm00}, and photoionization cross-sections are from \citet{vern96}. 
    }
    \label{fig:xray_spectrum}
\end{figure}

The X-ray spectrum of IC\,1297 was analyzed using the \texttt{XSPEC} spectral fitting package Version 12.14.1. 
The spectral fitting was restricted to the $0.3-2.0$~keV energy range because of the increasing calibration 
uncertainties at lower energies and the intrinsically soft nature of the source. The resulting X-ray spectra and best-fitting model are shown in Fig.~\ref{fig:xray_spectrum}.

We tested different grouping schemes with minimum counts per bin ranging from 1 to 20 and found that the resulting 
best-fit spectral parameters were consistent within uncertainties. A grouping criterion of 8 counts 
per bin has been adopted for the spectral analysis throughout this work. 
Given the limited photon statistics, both $\chi^2$ statistics and Cash statistics (C-stat) were explored. 
Since the source spectrum contains low-count bins, C-statistics provide a more appropriate treatment of 
Poisson-distributed data and were therefore adopted as the primary statistical estimator, while $\chi^2$ 
fitting was used as a consistency check.

The spectra were fitted using single- and two-temperature optically thin thermal plasma emission models, 
including \texttt{apec} and \texttt{vapec}, and a composite thermal plus non-thermal \texttt{apec+powerlaw} model. 
All tested models consistently indicate that the X-ray emission is dominated by a very soft thermal plasma. 
For the single-temperature models, the best-fit plasma temperature is in the range $kT \sim 0.08-0.10$ keV, 
corresponding to a characteristic plasma temperature $T \sim (0.9-1.2)\times10^{6}$ K. 

Allowing $N_{\mathrm H}$ to vary freely did not significantly improve the fits. 
In most cases, the $N_{\mathrm H}$ absorption column value of the best-fit model converged toward the imposed 
lower boundary, particularly for the \texttt{tbabs} models, indicating that the spectrum favors very low intrinsic 
absorption. Since the inferred plasma temperatures are low, much of the diagnostic spectral information is 
concentrated near the low-energy boundary at 0.3 keV of the adopted fitting range. Consequently, the absorption 
column density remains partially degenerate with the plasma temperature and instrumental calibration uncertainties. 
The HEASARC $N_{\mathrm H}$ tool, queried for IC\,1297 using the HI4PI map within a $0.1^\circ$ cone, returns an average Galactic H I column density of $6.71\times10^{20}$ cm$^{-2}$ (weighted average $6.66\times10^{20}$ cm$^{-2}$) \citep{HI4PI2016}. We regard this value as an upper limit. {We adopt the logarithmic extinction coefficient
$c(\mathrm{H}\beta)=0.07$ reported for IC\,1297 by
\citet{Milingo2002IIA}. Using the adopted conversion
$E(B-V)=c(\mathrm{H}\beta)/1.47$, we obtain
$E(B-V)\simeq0.048$~mag. The empirical relation
$N_{\mathrm H}/E(B-V)=5.8\times10^{21}$~cm$^{-2}$~mag$^{-1}$
\citep{Bohlin1978} then gives
$N_{\mathrm H}\simeq2.8\times10^{20}$~cm$^{-2}$.
We adopt this extinction-based estimate in the X-ray spectral fits.}

The \texttt{vapec} models were explored to investigate possible abundance variations in the X-ray-emitting plasma.
These could not be robustly constrained because of the limited photon statistics and strong degeneracy 
among abundance, temperature, and absorption at low energies. 
Two-temperature plasma models produced lower $\chi^2$ and C-stat values than the corresponding 
single-temperature models, but the statistical improvement is modest considering the increased number of 
free parameters. The additional hot component remained poorly constrained, spanning $kT_2 \sim 0.3-1.5$ keV, 
and its normalization remained very small relative to the dominant cool component. 

The non-thermal \texttt{powerlaw} component was explored to investigate the possible presence of 
unresolved hard X-ray emission, but again the fit did not statistically require such a component, whose 
parameters remained otherwise poorly constrained.

Given that the \ep\ FXT spectra of IC\,1297 do not provide compelling evidence for a multi-temperature plasma structure or an additional non-thermal component, we adopt the absorbed single-temperature model, \texttt{tbabs}$\times$\texttt{apec}, as the preferred spectral description of the source. 

The best-fit model has a plasma temperature of  $kT = 0.086$\,keV ($T\sim1.0\times10^{6}$ K) and 
a $N_{\mathrm{H}} = 2.9 \times 10^{20}$\,cm$^{-2}$, with a C-statistic of $22.5$ for $17$ dof 
(C-stat/d.o.f.\,$= 1.32$), indicating a statistically acceptable description of the data. 
Using the XSPEC \texttt{cflux} convolution model, the observed 
flux in the 0.3--2.0\,keV energy band is $f_{\mathrm{X}} = 9.66 \times10^{-14}$ erg\,cm$^{-2}$\,s$^{-1}$, 
which corresponds to an intrinsic X-ray flux $F_{\mathrm{X}} = 1.63 \times10^{-13}$ erg\,cm$^{-2}$\,s$^{-1}$, 
with the confidence interval computed using the XSPEC command \texttt{steppar}. Although the spectrum is soft, 
it is worth emphasizing that the low value of $N_\mathrm{H}$ implies a small correction of $\approx 1.7$ between observed and 
intrinsic X-ray fluxes.

\subsection{X-ray Luminosity}
\label{subsec:flux_calc}

The distance estimates for IC\,1297 merit discussion as they impact its intrinsic X-ray luminosity.  
The Gaia DR3 source associated with the CSPN has a Bayesian distance estimate of $3.7^{+3.0}_{-1.1}$ kpc \citep{gaiadr31,BailerJones2021}. This is a distance estimate for the star, whereas the following statistical and expansion-parallax distance estimates refer to the PN. 
Otherwise \citet{distance4908pc} reported a distance of 4.9 kpc from a new calibration of the scaling relation for Galactic PNe supported on statistical parallax and the use of Magellanic Cloud PNe. 
Building on this, \citet{distance4965pc} enhanced the methodology by adding systematic error handling and comparisons with alternative distance scales, for 
a consistent distance of $5.0\pm1.0$ kpc. 
This agrees with the distance of $5.5\pm1.1$ kpc provided by \citet{Aller1986} using the expansion parallax. However, \citet{FrewDistance} applied a well calibrated surface brightness-radius relation to derive a PN distance of $3.78\pm1.07$ kpc. These CSPN and PN distance estimates are consistent within their uncertainties.
We adopt the Frew et al. PN distance, rounded to 3.8 kpc, here.  
 
Therefore, at the most conservative distance of 3.8 kpc, the intrinsic X-ray flux in the 0.3--2.0 keV band of $1.63\times10^{-13}$ erg~cm$^{-2}$~s$^{-1}$ implies a luminosity of $L_{\mathrm X}=(2.8^{+2.1}_{-1.5})\times10^{32}$ erg~s$^{-1}$. 

This places IC\,1297 in the same X-ray luminosity range as BD+30$^\circ$3639 \citep{Joel2000BD} and NGC\,5189 \citep{ngc5189xray}, the previous most 
luminous known XPNe. 
BD+30$^\circ$3639 and NGC\,5189 also have [WR] CSPNe and both 
share morphological similarities to IC\,1297.

\subsection{Spatial analysis}
\label{subsec:xray_symmetry}

We analysed the X-ray spatial symmetry of IC\,1297 using a merged \ep\ FXT event list centred at its CSPN position. 
The instrument HPD of \qty{22}{arcsec} sets the 
effective resolution, and we adopt two reference radii from the optical structure: $r_{\mathrm{neb}} = \qty{5.4}{\arcsec}$ 
(nebular rim) and $r_{\mathrm{halo}} = \qty{19}{\arcsec}$ (outer halo). 
A local background of $\Sigma_{\mathrm{bg}} = \qty{1.54e-2}{counts\,arcsec^{-2}}$ 
was measured from the $100\arcsec$--$150\arcsec$ annulus.

\begin{figure}[h]
    \centering
    \includegraphics[width=0.4\columnwidth]{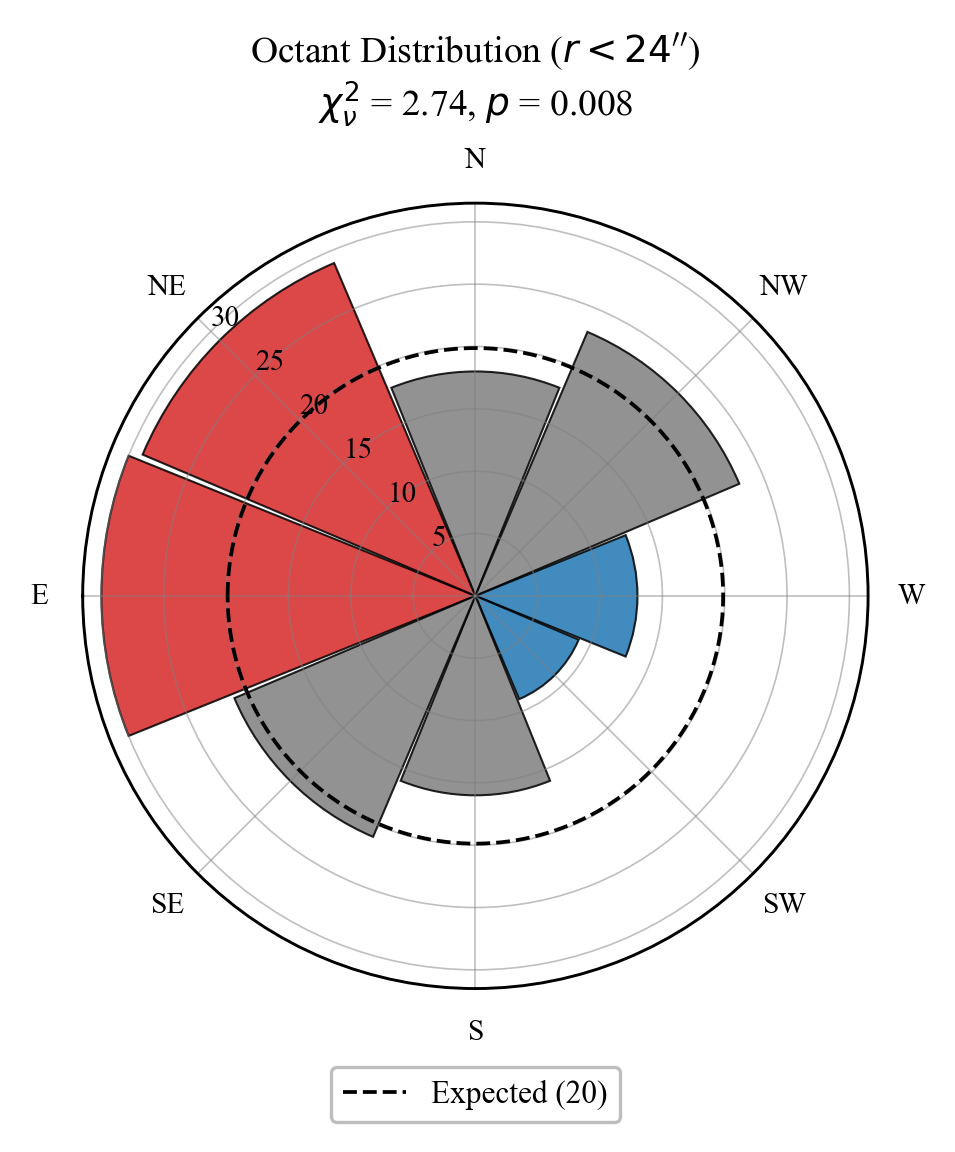}
    \caption{%
        Octant photon-count polar histogram within $r<24''$
        (north up, east left).
        {Bars are red ($N>1.25\,\bar{N}$), blue
        ($N<0.75\,\bar{N}$), or grey
        ($0.75\,\bar{N}\leq N\leq1.25\,\bar{N}$).
        Here, $N$ denotes the observed count in an octant, and
        $\bar{N}=19.9$ is the mean count per octant, used as the
        expected count under isotropy. The dashed circle marks
        this isotropic expectation
        ($\chi^2_\nu=2.74$, 7~dof, $p=0.008$).} 
        }
    \label{fig:symmetry}
\end{figure}

The flux-weighted centroid within the HPD aperture is offset from the optical nucleus by \qty{2.6}{arcsec} along $\mathrm{PA} \approx \ang{11}$, exceeding the $\lesssim 1''$ FXT astrometric uncertainty and 
indicating a genuine displacement of the X-ray emission.
To test azimuthal symmetry we applied $\chi^2$ tests for uniformity within the HPD aperture ($r < 24''$), 
reporting the reduced statistic $\chi^2_\nu \equiv \chi^2/\nu$ in Fig.\,\ref{fig:symmetry}.  
The east half contains 99 counts versus 60 in the west half ($\chi^2_\nu = 9.57$, 1\,dof, $p = 0.002$). 
With 93 counts, the north half similarly exceeds the 66 counts in the southern half 
($\chi^2_\nu = 4.58$, $p = 0.032$). 

Dividing the aperture into eight equal sectors (Fig.\,\ref{fig:symmetry}) yields counts of $30$ (E), $29$ (NE), $18$ (N), $23$ (NW), $13$ (W), $9$ (SW), $16$ (S), and $21$ (SE) against an isotropic expectation of $\bar{N}_{\mathrm{sec}} = 19.9$, giving $\chi^2_\nu = 2.74$ (7\,dof, $p = 0.008$).
The second-moment tensor of the photon distribution confirms an elliptical morphology ($e = 0.13$, $b/a = 0.87$).

The consistent east-to-north-east excess seen in all tests points to an intrinsic asymmetry of the hot gas distribution, 
plausibly driven by an anisotropic stellar wind interacting with a non-uniform circumstellar envelope \citep[e.g.][]
{ChanPlaNS1, ChanPlaNS2}, by confinement against a density gradient in the surrounding ISM \citep{Wareing2007}, or by internal extinction variations, as in the case of BD+30$^\circ$3639 \citep{Kastner2002Asymmetries}. 
Higher-resolution X-ray observations would be needed to distinguish between these scenarios.

\section{Discussion}
\label{sec:discussion}

\subsection{Nature of the X-ray Emission}
\label{subsec:nature_emission}
 
The sensitivity and spectral resolution of \ep\ FXT allows detecting and excising several background sources 
in the vicinity of IC\,1297 that were not resolved in the previous \erosita\ detection. Otherwise, the 
morphology of the \erosita\ and \ep\ X-ray images are broadly consistent, given the very small count 
number of the \erosita\ detection and the different energy responses and point spread functions of the 
two instruments. Given the spatial resolution of the present X-ray observations,
the emission can be interpreted to be enclosed within the central region of IC\,1297. 
This spatial distribution is consistent with the ISW model, where the fast 
wind from the central star shocks against the previously ejected slow AGB wind, creating a hot interior bubble \citep{Kwok1978}.

The spectral properties are indicative of a thermal plasma at an X-ray temperature 
$T\sim1.0\times10^{6}$~K ($kT \sim 0.086$ keV).  This is also consistent with expectations for shocked 
stellar wind bubbles in PNe \citep{Steffen2008}, as confirmed with X-ray observations \citep[see, for instance, the compilation by][]{Ruiz2013}.
The X-ray luminosity of IC\,1297 of $L_{\mathrm{X}} \sim 2.8 \times 10^{32}$ erg~s$^{-1}$ is comparable 
to that of other XPNe such as NGC\,6543 \citep[$L_{\mathrm{X}} \sim 10^{31}$ erg~s$^{-1}$,][]{Chu2001} and 
BD+30$^\circ$3639 \citep[$L_{\mathrm{X}} \sim 1.6 \times 10^{32}$ erg~s$^{-1}$][]{Joel2000BD}.

IC\,1297 hosts a [WR] CSPN of type [WC3]. 
Its terminal wind velocity is estimated to be in the range of $-3230$ to $-3670$~km~s$^{-1}$ based on P~Cygni 
profiles of the O\,\textsc{v} $\lambda 1371$ and C\,\textsc{iv} $\lambda 1549$ UV lines \citep{Feibelman1998}. 
This terminal wind velocity is at the upper high-velocity range for [WR] CSPNe and consistent with the shock 
thermalization required to produce the observed X-ray temperatures. Indeed, the high-dispersion optical spectra reported by \citet{Aller1986} show a complex ionization structure in IC\,1297, with evidence for shocks from line width measurements. 

We conclude that the spatial and spectral properties of IC\,1297's X-ray emission indicate that it arises from a hot bubble in the PN's innermost regions. This X-ray emission is powered by the fast stellar wind from its [WC3] CSPN and is consistent with the shock signatures inferred from the optical line widths reported by \citet{Aller1986}.

\subsection{Comprehensive investigation of XPNe with H-rich and H-poor CSPNe}
\label{subsec:comparison}

Appendix~\ref{tab:xpne} compiles the X-ray properties of the 41 Galactic PNe currently with X-ray detections taken from the ChanPlaNS project \citep{ChanPlaNS1,ChanPlaNS2, ChanPlaNS3}, and
individual object studies. 
For each we list the observatory, the observed $f_{\rm X}^{\rm obs}$ and intrinsic $F_{\rm X}^{\rm int}$ X-ray fluxes, the adopted distance $D$ \citep[taken 
as $D_{\rm mean}$ from][with its associated $1\sigma$ uncertainty]{FrewDistance}, and the resulting X-ray luminosity derived as 
$L_{\rm X} = 4\pi D^2 F_{\rm X}^{\rm int}$. 
Additional information on the CSPN spectral classification and on the spatial properties of the X-ray emission, whether it 
corresponds to diffuse emission (D) or to a point-source (P), has been added.

\begin{figure}[h]
\centering
\includegraphics[width=0.7\textwidth]{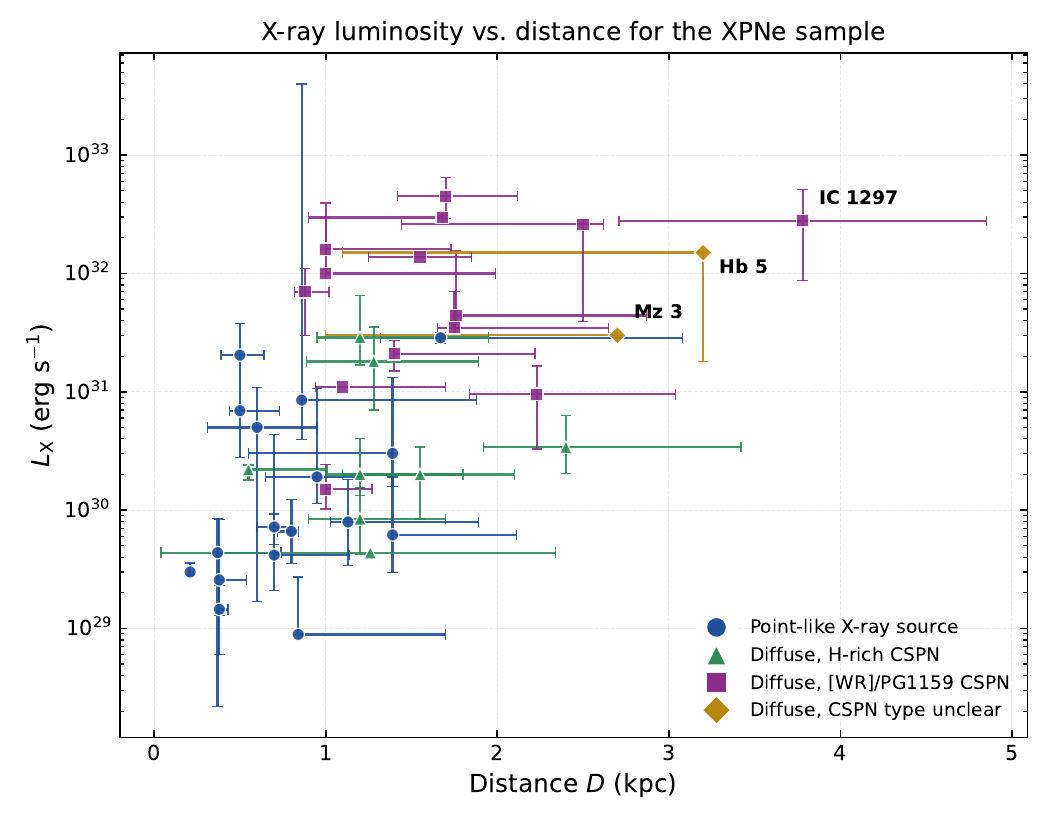}

\caption{
Comparison of distance and X-ray luminosity of IC\,1297 with other known XPNe in a luminosity 
versus distance plot. IC\,1297 is more distant than all other previously X-ray detected PNe by a 
factor of $\sim1.5$ and, within the uncertainties, among the most X-ray-luminous known PNe. 
}
\label{fig:comparison}
\end{figure}

The information in Appendix~\ref{tab:xpne} has been used to plot $L_{\rm X}$ as a function of $D$ for
the full sample of XPNe in Figure~\ref{fig:comparison}. 
Error bars on $D$ correspond to the $1\sigma$ distance uncertainties
from \citep{FrewDistance}, whereas those on $L_{\rm X}$ are propagated
from the distance (via $L_{\rm X}\propto D^2$) and $F_\mathrm{X}^\mathrm{int}$
uncertainties when available. For sources whose publication did not
report an uncertainty range for $L_{\rm X}$, we derived one ourselves
by propagating the reported $F_{\rm X}^{\rm int}$ and distance
uncertainties in quadrature via
$\sigma_{L}/L=\sqrt{(\sigma_F/F)^2+(2\sigma_D/D)^2}$, applied
separately to the upper and lower uncertainties. For Hb~5, this linear
approximation would formally give a negative $L_{\rm X}$ owing to the
large relative distance uncertainty, so the lower bound was instead
evaluated directly from $L=4\pi D^2 F$ at the lower distance limit.
Figure~\ref{fig:comparison} displays, for the first time, a clear correlation between 
distance and X-ray luminosity, reflecting the observational bias towards detecting only the brightest 
XPNe at large distances.  

The X-ray emission from CSPN is, in general, fainter than that from diffuse 
sources. 
The closer and less luminous XPNe in Figure~\ref{fig:comparison} are dominated by point source X-ray emitters, most with X-ray luminosities in the range $\approx 4\times10^{29}$ 
to $\approx 2\times10^{30}$ erg~s$^{-1}$, with the exception of NGC\,6337 and the very notable 
case of Hb\,5 and Mz\,3, which are possibly symbiotic stars according to HASH. 
The median luminosity of XPNe point-sources is $\approx 7\times10^{29}$ erg~s$^{-1}$. 

XPNe with diffuse emission are brighter and seem to belong to two different families of hot bubbles 
powered by H-rich and [WR] CSPNe.  The former are less frequent (36\% of XPNe with diffuse emission). 
They have X-ray luminosities in the range from a few times $10^{29}$ to a few times $10^{31}$ erg~s$^{-1}$, 
with a median X-ray luminosity $\approx 2\times10^{30}$ erg~s$^{-1}$, i.e., they are generally brighter than 
the XPN point-sources. The PNe with a [WR] CSPN have the highest X-ray luminosities, reaching up to 
several times $10^{32}$ erg~s$^{-1}$, with a median X-ray luminosity $\approx 9\times10^{31}$ erg~s$^{-1}$. 

The relatively high X-ray luminosities of XPNe with [WR] CSPNe have been noted previously \citep{Kastner2008NGC5315}. In our sample, such objects account for 64\% of XPNe with diffuse emission; this trend is shown clearly in Figure~\ref{fig:comparison}.
Indeed, the top ten most luminous XPNe in this plot all have [WR]/PG1159 CSPNe and diffuse X-ray emission.  
This is attributed to the stronger stellar winds of [WR] CSPNe. Recent radiation-hydrodynamical simulations of these sources 

\citep{OrozcoDuarte2026} show that their C- and O-rich stellar winds affect the production of X-rays inside their hot bubbles. 
Accordingly, [WR] CSPN produce more energetic hot bubbles compared to their H-rich counterparts due to their higher stellar wind mechanical luminosities and the strong radiative cooling of C-rich plasmas. 
These simulations predict X-ray luminosities up to $2\times10^{32}$ erg~s$^{-1}$ for typical [WR] PNe, at least 
one order of magnitude higher than hot bubbles powered by H-rich central stars. 
These predictions are fully consistent with the intrinsic X-ray luminosity of the brightest XPNe with [WR] CSPNe BD+30$^\circ$3639, IC\,1297, and NGC\,5189 ($\approx 3\times10^{32}$ erg~s$^{-1}$), and with the ratio between the median X-ray luminosity of XPNe with H-rich ($\approx 2\times10^{30}$ erg~s$^{-1}$) and [WR]-type ($\approx 9\times10^{31}$ erg~s$^{-1}$) CSPNe.

\section{Conclusions}
\label{sec:conclusions}

This work presents the first \ep\ FXT X-ray detection of a PN, namely IC\,1297. 
It exhibits soft, diffuse centrally enclosed X-ray emission, with a plasma temperature 
of $kT = 0.086$ keV ($T_\mathrm{X} \sim 1.0\times10^6$~K). The absorption-corrected X-ray luminosity in 
the 0.3--2.0 keV band is $L_{\mathrm{X}} = 2.8 \times 10^{32}$ erg s$^{-1}$. The spatial and spectral 
properties and the X-ray luminosity from IC\,1297 are consistent with those 
expected for hot bubbles in PNe. 
The X-ray luminosity of IC\,1297 makes it one of the most luminous known XPNe.  We also present a comprehensive compilation of X-ray detections of all currently known XPNe. 
As XPNe are faint X-ray sources, with observed X-ray fluxes in the range $\approx 10^{-15} - 6\times10^{-13}$ erg~cm$^{-2}$~s$^{-1}$, and their emission is soft, there is a bias towards detecting the nearest sources, which results in a correlation between X-ray luminosity and distance. 
Point-sources are typically the faintest, whereas there is a clear segregation between XPN with hot bubbles powered by H-rich and [WR] CSPNe, the former being less frequent and fainter than the latter. 
These comparisons show clearly for the first time that the ten most luminous XPNe all have hot bubbles powered by [WR] or PG\,1159 CSPNe.
Figure~\ref{fig:comparison} elaborates on previous X-ray observations showing that PNe with [WR]-type CSPNe tend to host the most luminous hot bubbles \citep{Kastner2008NGC5315}. This trend is consistent with models in which wind-wind collisions produce PN hot bubbles \citep{Kwok1978,OrozcoDuarte2026}; the stronger, C-rich winds of [WR] CSPNe can yield higher X-ray luminosities.

At the adopted distance of 3.8~Kpc, we establish IC\,1297 as the most distant and, within the uncertainties,
among the most X-ray-luminous known XPNe. It extends the known distance horizon for hot-bubble X-ray 
detections by a factor of $\sim1.54$. This opens a new observational window on hot bubble formation and evolution 
for Galactic PNe that would benefit from further X-ray observations. 
Targeted \ep\ observations can provide an extended distance baseline for future multi-wavelength studies of 
shocked gas in evolved stellar systems, given its proven capability to discover and characterize faint X-ray 
sources in the Galactic plane. These observations would be limited, however, by the spatial and spectral 
resolution of the \ep\ FXT instruments. Future observations with proposed facilities offering higher spectral 
resolution (e.g., XRISM and Athena) would enable measurement of metal abundances through line diagnostics, 
detection of possible non-equilibrium ionization effects, spatial mapping of temperature variations within 
the nebula, as well as searches for hard X-ray emission from their CSPNe.

\begin{acknowledgments}

This work is based on the data obtained with the Einstein Probe, a space mission supported by the Strategic Priority Program on 
Space Science of the Chinese Academy of Sciences, in collaboration with the European Space Agency, the Max-Planck-Institute for 
Extraterrestrial Physics (Germany), and the Centre National d'Etudes Spatiales (France). We thank the 
Einstein Probe team for their efforts in instrument development, calibration, and data processing. We thank Yifan Sun for helpful discussions during the data processing.
Q.A.P.\ thanks the Hong Kong Research Grants Council for GRF research support under grants 17304024, 17326116, and 17300417.
M.A.G.\ acknowledges financial support from the Severo Ochoa grant CEX2021-001131-S funded by MCIN/AEI/10.13039/501100011033 and the MCIU grant PID2024-162229NB-I00 cofunded with FEDER funds.
This work made use of the University of Hong Kong/Australian Astronomical Observatory/Strasbourg Observatory H-alpha 
Planetary Nebula (HASH PN) database, hosted by the Laboratory for Space Research at the University of Hong Kong.
This work has made use of data from the European Space Agency (ESA) mission {\it Gaia} (\url{https://www.cosmos.esa.int/gaia}), 
processed by the {\it Gaia} Data Processing and Analysis Consortium (DPAC, 
\url{https://www.cosmos.esa.int/web/gaia/dpac/consortium}). Funding for the DPAC has been provided by national institutions, in 
particular the institutions participating in the {\it Gaia} Multilateral Agreement.
 We thank Anthropic's Claude for assistance with code debugging and language editing during manuscript preparation. All AI-assisted content was reviewed and verified by the authors.

\end{acknowledgments}

\software{
astropy \citep{Astropy2013, Astropy2018, Astropy2022}, 
XSPEC \citep{Arnaud1996}, 
SAOImage DS9 \citep{Joye2003},
matplotlib \citep{Hunter2007},
numpy \citep{Harris2020},
scipy \citep{Virtanen2020},
FXTDAS 1.30 \citep{epfxt_fxtdas_2024},
Claude Sonnet5\citep{Anthropic2026}
}

\appendix
\section{Compilation of X-ray-emitting planetary nebulae}
\label{app:xpne}

Table~\ref{tab:xpne} compiles the X-ray properties of the 41 Galactic planetary nebulae currently known to have X-ray detections. For each object, it lists the observatory, the observed and intrinsic X-ray fluxes, the adopted distance, the resulting X-ray luminosity, the CSPN spectral type, and whether the X-ray emission is diffuse (D), point-like (P), or both.

The fluxes and luminosities are taken from the source publications listed in the table notes and reference list below the table. The adopted PN distances and their $1\sigma$ uncertainties are based on the distance estimates in Tables~3 and 4 of \citet{FrewDistance}; where available, the central distance value is that adopted in the corresponding simulations. The X-ray luminosities are calculated as $L_{\rm X}=4\pi D^2F_{\rm X}^{\rm int}$. Notes g and h compare the adopted distances with independent published distance estimates and describe how the reported uncertainty bounds were treated.

\begin{longrotatetable}
\begin{longtable}{lccccccc}

\caption{X-ray properties of Galactic planetary nebulae with X-ray detections.\label{tab:xpne}} \\

\hline
\multicolumn{1}{l}{PN} &
\multicolumn{1}{c}{Observatory} &
\multicolumn{1}{c}{$f_{\rm X}^{\rm obs}$} &
\multicolumn{1}{c}{$F_{\rm X}^{\rm int}$} &
\multicolumn{1}{c}{Distance} &
\multicolumn{1}{c}{$L_{\rm X}$}&
\multicolumn{1}{c}{CSPNe Type}&
\multicolumn{1}{c}{X ray type}
\\
& &
\multicolumn{1}{c}{(erg cm$^{-2}$ s$^{-1}$)} &
\multicolumn{1}{c}{(erg cm$^{-2}$ s$^{-1}$)} &
\multicolumn{1}{c}{(kpc)} &
\multicolumn{1}{c}{(erg s$^{-1}$)} \\
\hline
\endfirsthead

\multicolumn{8}{c}{{\tablename\ \thetable{} -- continued}} \\
\hline
\multicolumn{1}{l}{PN} &
\multicolumn{1}{c}{Observatory} &
\multicolumn{1}{c}{$F_{\rm X}^{\rm obs}$} &
\multicolumn{1}{c}{$F_{\rm X}^{\rm int}$} &
\multicolumn{1}{c}{Distance} &
\multicolumn{1}{c}{$L_{\rm X}$} &
\multicolumn{1}{c}{CSPNe Type}&
\multicolumn{1}{c}{X ray type}\\
& &
\multicolumn{1}{c}{(erg cm$^{-2}$ s$^{-1}$)} &
\multicolumn{1}{c}{(erg cm$^{-2}$ s$^{-1}$)} &
\multicolumn{1}{c}{(kpc)} &
\multicolumn{1}{c}{(erg s$^{-1}$)} \\
\hline
\endhead

\hline
\endfoot

A30\tablenotemark{a}               & Chandra + XMM-Newton & $1.00^{+0.31}_{-0.20}\times10^{-13}$  & $1.19\times10^{-13}$\tablenotemark{c} & $1.76^{+1.11}_{-0.00}$\tablenotemark{h}  & $4.4^{+11.15}_{-0.00}\times10^{31}$\tablenotemark{m}&[WC5]& D,P \\ 
A78\tablenotemark{b}             & XMM-Newton           & $(2.2\pm0.8)\times10^{-14}$  & $8.95\times10^{-14}$\tablenotemark{c} & $1.40^{+0.82}_{-0.00}$\tablenotemark{h} & $(2.1\pm0.6)\times10^{31}$ &[WC]&D,P\\ 
BD$+30^\circ3639$ & Chandra              & $(5.7\pm0.68)\times10^{-13}$  & $1.34\times10^{-12}$\tablenotemark{c} & $1.00^{+0.73}_{-0.00}$\tablenotemark{h} & $1.6^{+2.34}_{-0.00}\times10^{32}$\tablenotemark{m}\tablenotemark{i} & [WC9] & D \\
DS\,1             & Chandra              & $9.5\times10^{-15}$  & $1.2\times10^{-14}$ & $0.70^{+0.10}_{-0.10}$ \tablenotemark{g} & $7.2^{+2.06}_{-2.06}\times10^{29}$\tablenotemark{m}&sdO&P \\
Hb\,5             & XMM-Newton+ Chandra  & $7.5\times10^{-15}$ & $1.2\times10^{-13}$  & $3.2^{+0.0}_{-2.1}$\tablenotemark{h} & $1.5^{+0.00}_{-1.32}\times10^{32}$\tablenotemark{m} & SySt? / binary &D\\
HbDs\,1           & Chandra              & $4.79^{+0.58}_{-0.62}\times10^{-15}$  & $8.71^{+1.76}_{-1.47}\times10^{-15}$ & $0.80^{+0.04}_{-0.08}$\tablenotemark{g}  & $6.61^{+5.70}_{-3.06}\times10^{29}$&O(H)&P \\
HFG\,1            & Chandra              & $5.4\times10^{-14}$ & $1.1\times10^{-13}$ & $0.60^{+0.35}_{-0.29}$ \tablenotemark{g}& $5.0^{+5.83}_{-4.83}\times10^{30}$\tablenotemark{m} &sdO&P\\
IC\,418           & Chandra              & $2.5\times10^{-15}$  & $4.88\times10^{-15}$ & $1.2^{+0.5}_{-0.3}$\tablenotemark{g}& $8.4^{+7.0}_{-4.2}\times10^{29}$\tablenotemark{m}  &Of(H)&D\\
IC\,1297 & eROSITA + EP         & $1.0\times10^{-13}$  & $1.6^{+1.0}_{-0.6}\times10^{-13}$ & $3.8 \pm 1.1$\tablenotemark{g}& $2.8^{+2.3}_{-1.9}\times10^{32}$ &[WC3]& D \\
IC\,4593      & Chandra              & $3.3\times10^{-15}$  & $5.0\times10^{-15}$ &  $2.4^{+1.0}_{-0.5}$\tablenotemark{g}& $3.4^{+2.9}_{-1.4}\times10^{30}$\tablenotemark{m} &Of(H)& D,P\\
K\,1-16           & Chandra + XMM-Newton & $(6.6\pm0.9)\times10^{-14}$& $(8.6\pm0.9)\times10^{-14}$  &  $1.7^{+1.4}_{-0.4}$\tablenotemark{g}& $(2.9\pm0.3)\times10^{31}$  &PG1159&P\\
Lo\,16\tablenotemark{i}         & Chandra              & $8.31^{+5.17}_{-2.56}\times10^{-16}$  & $1.047\times10^{-15}$ & $0.84^{+0.86}_{-0.00}$\tablenotemark{h} & $8.91^{+18.24}_{-0.00}\times10^{28}$\tablenotemark{m} &O(H)3&P\\
LoTr\,5& Chandra + XMM-Newton & \ldots & $2.3^{+0.0}_{-1.4}\times10^{-13}$ &  $0.50^{+0.23}_{-0.06}$ \tablenotemark{g}& $6.91\times10^{30}$&G5III-IV&P \\
Mz\,3\tablenotemark{d}  & Chandra              & $7.0\times10^{-15}$  & $3.44\times10^{-14}$\tablenotemark{c}  & $2.7^{+0.0}_{-1.7}$ \tablenotemark{g}& $3\times10^{31}$ &SySt?/[WC]& D,P\\
NGC\,1360\tablenotemark{i}        & Chandra              & $1.45^{+0.79}_{-0.35}\times10^{-14}$ & $1.479^{+2.787}_{-0.432}\times10^{-14}$ & $0.38^{+0.16}_{-0.00}$\tablenotemark{g}& $2.570^{+5.748}_{-1.282}\times10^{29}$ &sdO & P \\
NGC\,1514         & Chandra              & $2.818^{+0.344}_{-0.364}\times10^{-15}$ & $1.41^{+8.77}_{-0.15}\times10^{-13}$ & $0.37^{+0.37}_{-0.00}$\tablenotemark{h} & $4.37\pm4.15\times10^{29}$ &sdO&P\\
NGC\,40        & Chandra              & \ldots & $(1.3\pm0.42)\times10^{-14}$  & $1.00^{+0.27}_{-0.00}$\tablenotemark{h} & $1.5^{+0.94}_{-0.48}\times10^{30}$\tablenotemark{m}  &[WC8]&D\\
NGC\,246\tablenotemark{i,l} & Chandra              & $5.25^{+0.38}_{-0.35}\times10^{-13}$  & $6.92^{+1.40}_{-1.16}\times10^{-13}$ & $0.50^{+0.14}_{-0.11}$\tablenotemark{g} & $(2.04 \pm 1.76) \times 10^{31}$  &PG1159&P\\
NGC\,2371\tablenotemark{i,l} & Chandra              & $1.23^{+0.68}_{-0.47}\times10^{-13}$ & $1.45 \times 10^{-13}$ & $1.75^{+0.90}_{-0.10}$\tablenotemark{g}& $3.47^{+3.57}_{-0.40}\times10^{31}$\tablenotemark{m} &[WO1]&D,P\\%
NGC\,2392\tablenotemark{e}       & Chandra              & $3.9\times10^{-14}$  & $9.18\times10^{-14}$ & $1.28^{+0.61}_{-0.39}$\tablenotemark{g} & $1.8^{+1.72}_{-1.10}\times10^{31}$\tablenotemark{m}  &O6f&D,P\\
NGC\,2867\tablenotemark{n}         & eROSITA              & $1.60^{+0.75}_{-0.53}\times10^{-14}$  & \ldots & $2.23^{+0.81}_{-0.39}$ \tablenotemark{g}& $9.52^{+6.98}_{-6.24}\times10^{30}$&[WO2]&D\\
NGC\,3242         & XMM-Newton       & \ldots    & $4.20^{+0.70}_{-1.10}\times10^{-14}$   &  $0.55^{+0.46}_{-0.00}$\tablenotemark{g} & $2.2^{+0.2}_{-0.4}\times10^{30}$ &O(H)&D\\
NGC\,3918\tablenotemark{j} & Chandra              & $5.01 \times 10^{-15}$ & $6.96\times10^{-15}$\tablenotemark{c} & $1.55^{+0.55}_{-0.45}$\tablenotemark{g} & $2.0^{+1.42}_{-1.16}\times10^{30}$\tablenotemark{m} &\ldots & D\\
NGC\,4361\tablenotemark{j} & Chandra              & $1.78^{+2.39}_{-1.45} \times 10^{-14}$  & $ 1.78^{+7.14}_{-0.66} \times 10^{-14}$ & $0.95^{+0.26}_{-0.30}$\tablenotemark{g}  & $ 1.91^{+8.81}_{-0.78} \times 10^{30}$ &[WC]& P\\
NGC\,5189
& XMM-Newton           & $(6.8\pm2.0)\times10^{-14}$  & $(8.87\pm0.52)\times10^{-13}$ & $1.7^{+0.0}_{-0.8}$\tablenotemark{h} & $(2.98\pm0.12)\times10^{32}$&[WO1]& D \\
NGC\,5315         & Chandra              &   \ldots &$3.5\times10^{-13}$ & $2.5^{+0.12}_{-1.06}$\tablenotemark{g} & $2.6^{+0.25}_{-2.21}\times10^{32}$\tablenotemark{m}  &[WC4]&D\\
NGC\,6153 \tablenotemark{j}        & Chandra              & $ 7.59 \times 10^{-14}$  & \ldots &$1.10^{+0.60}_{-0.16}$\tablenotemark{g}& $ 1.10 \times 10^{31}$  &[WC]-PG1159&D\\
NGC\,6337         & Chandra              & $8.13^{+1.87}_{-2.63}\times10^{-15}$  & $9.55^{+4256.25}_{-3.52}\times10^{-14}$ &$0.86^{+1.02}_{-0.00}$\tablenotemark{h} & $8.51^{+3972.56}_{-4.53}\times10^{30}$ &Wels & P\\
NGC\,6369\tablenotemark{g,j}& Chandra              & $4.90 \times 10^{-13}$ & \ldots & $1.55^{+0.30}_{-0.30}$  \tablenotemark{g}& $1.38 \times 10^{32}$ &[WO1]&D \\
NGC\,6445 \tablenotemark{i}        & Chandra              & $4.27^{+1.36}_{-1.31}\times10^{-15}$ & $1.32^{+3.81}_{-0.29}\times10^{-14}$ & $1.39^{+0.00}_{-0.84}$\tablenotemark{h} & $3.02^{+10.16}_{-1.44}\times10^{30}$  & \ldots&P\\
NGC\,6543         & Chandra              & $8\times10^{-14}$  & $8\times10^{-13}$ & $1.00^{+0.99}_{-0.00}$\tablenotemark{h}  & $1.0^{+1.98}_{-0.00}\times10^{32}$\tablenotemark{m} &[WC]&D,P \\
NGC\,6826      & Chandra              & $9.0\times10^{-15}$  & $9.9\times10^{-15}$ & $1.20^{+0.60}_{-0.20}$\tablenotemark{g} & $2.0^{+2.0}_{-0.67}\times10^{30}$\tablenotemark{m}  &O3f&D,P\\ 
NGC\,6853\tablenotemark{k}         & Chandra              & $(1.26\pm0.10)\times10^{-14}$ & \ldots &  $0.38^{+0.05}_{-0.00}$\tablenotemark{h} & $(1.45\pm0.85)\times10^{29}$ &DAO&P \\
NGC\,7008\tablenotemark{p,i,l} & Chandra              & $3.63^{+1.74}_{-1.06}\times10^{-15}$  & $7.08^{+62.10}_{-1.95}\times10^{-15}$ & $0.70^{+0.44}_{-0.00}$\tablenotemark{h} & $4.17^{+39.48}_{-2.08}\times10^{29}$  &O7&P\\
NGC\,7009         & XMM-Newton           & $7.2\times10^{-14}$  & $1.7\times10^{-13}$ & $1.2^{+0.75}_{-0.25}$\tablenotemark{g}& $2.88^{+3.60}_{-1.20}\times10^{31}$\tablenotemark{m} &O(H)& D,P\\ 
NGC\,7026         & XMM-Newton           & $(8.8\pm1.2)\times10^{-15}$  & $1.3^{+0.5}_{-0.4}\times10^{-12}$ & $1.7^{+0.42}_{-0.28}$\tablenotemark{g} & $(4.5^{+2.0}_{-1.6})\times10^{32}$&[WC3]& D  \\
NGC\,7027         & Chandra              & \ldots  & $6.95\pm 4.75\times10^{-13}$ & $0.88^{+0.14}_{-0.06}$ \tablenotemark{g}& $7\pm 4\times10^{31}$ &[WR]&D\\
NGC\,7094\tablenotemark{l} & Chandra              & $2.29^{+1.02}_{-0.81}\times10^{-15}$ & $2.63^{+4.13}_{-0.89}\times10^{-15}$ & $1.39^{+0.72}_{-0.00}$ \tablenotemark{g}& $6.17^{+12.89}_{-3.21}\times10^{29}$ &sdB&P\\
NGC\,7293          & Chandra              & \ldots & $5.68\times10^{-14}$ & $0.21^{+0.02}_{-0.006}$\tablenotemark{g} &$3.0^{+0.57}_{-0.17}\times10^{29}$\tablenotemark{m} &DAO& P\\
NGC\,7662 \tablenotemark{j}        & Chandra              & $2.82\times10^{-15}$ & \ldots & $1.26^{+1.08}_{-1.22}$ \tablenotemark{g}& $4.37 \times 10^{29}$ &sdO&D\\
Sp\,1  \tablenotemark{i}           & Chandra              & $1.23^{+0.32}_{-0.28}\times10^{-15}$ & $5.25^{+4.08}_{-2.36}\times10^{-15}$ & $1.13^{+0.76}_{-0.10}$\tablenotemark{g} & $7.94^{+10.25}_{-4.55}\times10^{29}$  &$O(H)_{3-5}$&P\\
\end{longtable}

\textit{Note.} 
$f_{\rm X}^{\rm obs}$ is the observed (absorbed) X-ray flux and $F_{\rm X}^{\rm int}$ is the intrinsic (unabsorbed) X-ray flux, both as reported in the cited source publications; ``\ldots'' indicates that the source publication did not separately report a flux.
$L_{\rm X}$ is the X-ray luminosity derived from the intrinsic flux and distance. 
X-ray fluxes and luminosities are computed over the energy band adopted in each source publication (typically a soft band, but the exact range varies --- e.g.\ 0.3--2.0~keV, 0.2--1.5~keV, up
to $\sim3.0$~keV in some cases --- depending on the instrument and spectral model used). 
No attempt was made to convert these to a common energy band, thus the source-to-source comparison of $F_{\rm X}$ or $L_{\rm X}$ carries this additional systematic uncertainty.

TS\,01 (PN\,G135.9+55.9), a Galactic halo object at $\sim21$~kpc, is not included because we consider its PN classification uncertain. Its unusual optical spectrum and original identification as a cataclysmic variable are discussed by \citet{Tovmassian2001}. Its central X-ray point source is interpreted as supersoft photospheric emission from the hot component of a double-degenerate binary \citep{Tovmassian2010}.

\tablenotetext{a}{A\,30 has both a diffuse and a point-source component \citep{a30Xray}. 
}

\tablenotetext{b}{A 78 has both a diffuse and a point-source component \citep{Toal2015}. 
}

\tablenotetext{c}{$F_{\rm X}^{\rm int}$ was not reported in the source publication and is instead derived here from the reported $L_{\rm X}$ and adopted distance via $F_{\rm X}^{\rm int} = L_{\rm X}/(4\pi d^2)$; no uncertainty is propagated.
}

\tablenotetext{d}{Mz\,3: its total intrinsic X-ray luminosity is $L_X \approx 3\times10^{31}\,(d/2.7\ \mathrm{kpc})^2\ \mathrm{erg\ s^{-1}}$, of which $\sim6\times10^{30}\,(d/2.7\ \mathrm{kpc})^2\ \mathrm{erg\ s^{-1}}$ arises from the central point source and $\sim2.4\times10^{31}\,(d/2.7\ \mathrm{kpc})^2\ \mathrm{erg\ s^{-1}}$ from the diffuse component \citep{Mz3kastner}.}

\tablenotetext{e}{NGC~2392: its CSPN has an observed flux of $1.3\times10^{-14}$ erg~cm$^{-2}$~s$^{-1}$ and a luminosity $9\times10^{30}$ erg~s$^{-1}$ \citep{Guerrero2019}.}

\tablenotetext{g}{The adopted distance lies inside the 1$\sigma$ range of the independent distance estimate\citep{FrewDistance}.}

\tablenotetext{h}{The adopted distance lies outside the
1$\sigma$ range of the independent distance estimate; the exceeded
bound is adopted as the limiting value on that side (uncertainty
clamped to 0), while the opposite-side uncertainty is computed
normally from the independent range\citep{FrewDistance}.
}

\tablenotetext{i}{\citet{ChanPlaNS3} adopts a model with fixed value of $N_\mathrm{H}$.}

\tablenotetext{j}{\citet{ChanPlaNS3} sets an upper limit for $T_\mathrm{X}$ of $1\times10^6$ K.}

\tablenotetext{k}{NGC\,6853 (M\,27) X-ray emission is outside the reliable ACIS calibration range, thus the values of $f_\mathrm{X}^\mathrm{obs}$ and $F_\mathrm{X}^\mathrm{int}$ should be regarded with caution. 
}

\tablenotetext{l}{\citet{ChanPlaNS3} indicates spectral fits are unreliable for NGC\,246, NGC\,2371-72, NGC\,7008, and NGC\,7094.}

\tablenotetext{m}{The uncertainty range for $L_{\rm X}$ was derived from the reported $F_{\rm X}^{\rm int}$ and distance uncertainties. 
}

\tablenotetext{n}{NGC\,2867: unlike most other entries in this table, the value of $f_{\rm X}^{\rm obs}$ does not come from a formal spectral fit, but it was derived from the eROSITA/eRASS1 net source counts and count rate converted to flux via an assumed energy-conversion factor \citep{Yuan_2025}}.

The compilation is based primarily on
\citet{ChanPlaNS1}, 
\citet{ChanPlaNS2}, 
\citet{ChanPlaNS3},
\citet{fluxXPNe}, 
\citet{ic418ngc2392ngc6826}, 
\citet{a30Xray}, 
\citet{Toal2015},
\citet{ic4593},
\citet{Liu2000}, 
\citet{Kastner2000},
\citet{Hb5XMMNewton}, 
\citet{MontezHFG1DS1LOTR5}, 
\citet{K116},
\citet{ngc40xray},
\citet{ngc1360xray1985},
\citet{ngc246xray},
\citet{Kastner_2008},
\citet{ngc2371xray},
\citet{Ruiz2011},
\citet{ngc4361},
\citet{Mz3kastner}, 
\citet{ngc5189xray}, 
\citet{Chu_2001ngc6543},
\citet{ngc7009xray},
\citet{xrayngc2346ngc7026},
\citet{Guerrero2001},
and \citet{ngc7027kastner}.

\end{longrotatetable}

\bibliography{sample701}
\bibliographystyle{aasjournalv7}

\end{document}